\documentclass[12pt,preprint]{aastex}
\def\({\left(}
\def\){\right)}
\def\[{\left[}
\def\]{\right]}
\begin{document}
\title {Spectra from a magnetic reconnection-heated corona
in AGN}
\author{ B. F. Liu, S. Mineshige, and K. Ohsuga}
\affil{Yukawa Institute for Theoretical Physics, Kyoto University,
Kyoto 606-8502, Japan}
\email{bfliu@yukawa.kyoto-u.ac.jp (BFL)}
\begin{abstract}
We investigate a  corona coupled with   
underlying disk through magnetic field and radiation field, and
present emergent spectra.
Due to buoyancy the magnetic flux loop emerges from the disk and
reconnects with other loops in the corona, thereby releasing the
magnetic energy to heat the coronal plasma. The energy is
then radiated away through Compton scattering. 
By studying the energy balance in the corona,  transition layer and disk, 
we determine the fraction ($f$) of accretion energy dissipated in the corona 
for given black-hole mass and  accretion rate, and
then determine the coronal and disk variables. 
 This allows us to calculate 
emergent spectra through Monte Carlo simulations. The
spectra are then determined as functions of black-hole
mass and accretion rate.   We find two types of solutions corresponding
for hard spectrum and soft spectrum. 
In the hard-spectrum solution, the accretion
energy is dominantly dissipated in the corona, supporting a strong
corona above a cool disk;  The hard X-ray spectral indices are the same for 
different accretion rate, 
i.e. $\alpha \sim 1.1$ ($F_\nu\propto \nu^{-\alpha}$).
 In the soft-spectrum 
solution, the accretion energy is mainly dissipated in the
disk. The
coronal temperature and density are quite low. Consequently, the spectra are
dominated by the disk radiation peaked
in UV and soft X-rays.
For low-luminosity systems ($L\la 0.2L_{\rm Edd}$),  there
exists only the solution of hard
spectra; While for high-luminosity systems ($L\ga 0.8L_{\rm Edd}$),
there exist both solutions of hard and soft spectra. 
For middle-luminosity systems ($0.2L_{\rm Edd}\la L\la 0.8L_{\rm Edd}$), 
besides the hard spectra,
moderately soft spectra composed of an inner soft-spectrum solution 
and an outer
hard-spectrum solution may occur, the softness of which increases with 
 increasing luminosity.
The hard spectra are
close to the observed spectra in Seyfert galaxies and radio-quiet QSOs.
The composite spectra may account for the diversity of broad band spectra
observed in narrow-line Seyfert 1 galaxies. 
\end{abstract}
\keywords {accretion, accretion disks--galaxies:
nuclei--X-rays:galaxies}

\section{Introduction}
It is commonly believed that the hard X-ray radiation of 
active galactic nuclei (AGNs) or galactic black hole candidates (GBHCs) 
arises from hot gas around accreting black holes,
either in forms of hot accretion flow or accretion-disk corona 
(e.g. Liang \& Nolan 1984; Mushotzsky, Done, \& Pounds 1993;
Narayan, Mahadevan, \& Quataert 1998).
The structure of hot accretion flows, such as advection-dominated
accretion flow (ADAF), has been extensively investigated  
 based on the vertically one-zone approximation (Kato, Fukue, \&
Mineshige 1998 and references therein).
The ADAF model (Ichimaru 1977; Narayan, Yi \& Mahadevan 1995; see also
Blandford \& Begelman 1999;
Igumenshchev \& Abramowicz 2000 for modifications) can reasonably fit  the
observed spectra of low luminosity AGNs and 
X-ray novae in quiescence.
Nevertheless, spectral features observed in ordinary AGNs, 
the big blue bump, the soft X-ray excess
and hard X-ray continuum,
and the 6.4 keV fluorescent iron lines, show strong observational
evidence for hot gas coexisting with cool gas in the vicinity of accreting black hole (e.g., Mushotzky et al. 1993),
which is presumably a corona lying above an accretion disk.
In disk-corona
models, the cold disk is embedded in the hot corona in a plane-parallel
slab. Except for the ion-illuminated disk (Deufel \& Spruit 2000;
Deufel, Dullemond, \& Spruit 2002), previous disk corona models for
AGNs or GBHCs 
(Haardt \& Maraschi
1991; 1993; Nakamura \& Osaki 1993; Svensson \& Zdziarski 1994; Dove et al. 1997; Kawaguchi,
Shimura, \& Mineshige 2001) need to assume a large fraction of accretion energy to be released
in the corona though the details of the coronal
heating mechanism remain unclear.    

Detailed investigation on the interaction between  an accretion
disk 
and a friction-heated corona  (Liu et al. 2002b) shows that
the energy deposit in the corona is not enough to keep itself hot
above the disk against strong Compton cooling in AGN systems. Other
heating besides the viscous one is required to maintain such a hot corona
lying above the disk and to produce the observed X-ray
luminosity. Comparison of the thermal energy
capacity in the corona and observed power of AGNs also shows  energy shortage
in the corona (Merloni \& Fabian 2001).  
One promising mechanism to heat the corona may be the magnetic
reconnection (e.g. Di Matteo 1998; Di
Matteo, Celotti, \& Fabian 1999; Miller \& Stone 2000; Machida,
Hayashi \& Matsumoto 2000) as a result of magnetorotational instabilities in
the disk and the buoyancy of the magnetic field (Tout \& Pringle 1992;
Miller \& Stone 2000). Despite of   the  complex radiation
and energy interaction between the disk main body and the corona,
magnetic fields  seem to play essential role in
producing time variations
and spatial inhomogeneity (e.g. Kawaguchi et al. 2000).

In a recent study (Liu, Mineshige, \& Shibata 2002a, hereafter Paper I), we show that, the magnetic reconnection can indeed
heat the corona to a temperature around $10^9$K and produce the
observed X-ray luminosity. 
In the present paper, we describe the  model  in a more
consistent way (Sect.\ref{s:model}), 
and then calculate the spectra from the corona and disk by
 Monte Carlo simulations. The computational results on the
coronal properties and emergent spectra are presented in
Sect.\ref{s:result}. Discussion on spectra  is
presented in Sect.\ref{s:discussion} and  our conclusions in Sect.\ref{s:conclusion}.
       
\section{The model}\label{s:model}

In our disk-corona model (Paper I), the disk is assumed to be a
classical Shakura \& Sunyaev (1973) disk.  The corona is a
plane-parallel corona which is tightly coupled with the underlying disk.
In the disk, magnetic fields are continuously
generated by dynamo action. Owing to the buoyancy 
magnetic flux loops emerge into the corona and reconnect with
other loops. Thereby, the
magnetic energy carried from the disk is released in the corona as thermal energy.
If the density of corona is not high enough, heat is conducted by electrons
from the corona to the chromosphere,  resulting in mass evaporation. 
Once the density of corona reaches a certain value,
Compton scattering becomes dominant in cooling, and eventually 
an equilibrium is established between the reconnection heating and Compton
cooling, a stationary corona is built up. The equations describing
these processes in the corona and at the
interface of disk and corona are (see Paper I),
\begin{equation}\label{e:energy}
{B^2\over 4\pi}V_A\approx {4kT\over m_ec^2}\tau^* cU_{\rm rad},
\end{equation} 
\begin{equation}\label{e:evap}
{k_0T^{7\over 2}\over \ell}\approx {\gamma\over \gamma-1} n k T {\(kT\over m_H\)}^{1\over 2}.
\end{equation}
In stead of $\tau$ in Paper I, we introduce effective optical depth,
 $\tau^*$, which   includes the
 isotropic incident photons undergoing up-scattering in a
 plane-parallel corona,
\begin{equation} 
\tau^*\equiv\lambda_\tau \tau=\lambda_\tau n\sigma_T\ell 
\end{equation}
with $\lambda_\tau$ being order of unit.

Eq.(\ref{e:energy}) and Eq.(\ref{e:evap}) determine the temperature
and density in the corona as function of energy densities of magnetic
field and radiation field.

The magnetic field strength $B$ is derived by assuming
 equipartition of gas energy and magnetic energy 
in the disk,
\begin{equation}\label{e:equipatition}
\beta\equiv {n_{\rm disk} k T_{\rm disk}\over {B^2\over
8\pi}}\sim 1.
\end{equation}
For a radiation pressure-dominated disk,
\begin{equation}\label{e:magr}
B=1.47\times 10^3 \alpha_{0.1}^{-{5\over 8}}\beta_1^{-{1\over 2}}m_8^{-{5\over 8}}\[\dot
m_{0.1}\phi\(1-f\)\]^{-1}r_{10}^{9\over 16} G;
\end{equation}
For a gas pressure-dominated disk,
\begin{equation}\label{e:magg}
B=7.18\times 10^4  \alpha_{0.1}^{-{9\over20}}\beta_1^{-{1\over 2}}m_8^{-{9\over 20}}\[\dot
m_{0.1}\phi\(1-f\)\]^{2\over 5}r_{10}^{-{51\over 40}} G,
\end{equation} 
where $f$ is the energy fraction dissipated in the corona;  
$\phi\equiv 1-\sqrt{R_*\over R}$ and $R_*$ is taken to
be the last stable orbit $3R_{\rm S}$;  $m_8$, $\dot m_{0.1}$, 
$r_{10}$, $\alpha_{0.1}$, $\beta_1$ are the mass of black hole, the
accretion rate, the
distance, the viscous coefficient, and the equipartition factor  in
units of $10^8M_\odot$, $0.1\dot M_{\rm Edd}$, $10R_{\rm S}$, 0.1,
and 1, respectively.

The energy density of soft photon field,  $U_{\rm rad}$, is contributed from both
intrinsic disk radiation $U_{\rm rad}^{\rm in}$ and reprocessed
radiation 
$U_{\rm rad}^{\rm re}$ of backward Compton emission from the corona  
\begin{equation}\label{e:uradin}
\begin{array}{ll}
U_{\rm rad}^{\rm in}& =aT_{\rm eff}^4=\frac{\displaystyle
4}{\displaystyle  c}\frac{\displaystyle 3G M\dot
M(1-f)\phi}{\displaystyle  8\pi
R^3}\\
& =1.14\times 10^5 m_8^{-1}\dot m_{0.1}\phi (1-f)r_{10}^{-3}{\rm ergs\ cm^{-3}},
\end{array}
\end{equation} 

\begin{equation}\label{e:uradre}
U_{\rm rad}^{\rm re}=0.4\lambda_u U_{\rm B}.
\end{equation} 
In Eq.(\ref{e:uradre}), a new factor $\lambda_u$ is added to the
evaluation of seed field in Haardt \& Maraschi (1991; 1993), which  includes two
effects on the soft photon energy, i.e. the dependence on the coronal
temperature and optical depth, and the ratio of Alfv\'en speed to
the light speed since the magnetic reconnection releases energy 
at Alfv\'en speed while this energy is radiated away at light speed. 
 
Now we are allowed to derive  two sets of coronal
solutions,  corresponding to
a radiation pressure-dominated disk and a gas pressure-dominated disk,
respectively.  
 From Eqs.(\ref{e:energy}), (\ref{e:evap}), (\ref{e:magr}), and  (\ref{e:uradin}), we derive a coronal solution above a radiation pressure-dominated disk, 
\begin{equation}\label{e:Tr1}
T_1=1.36\times 10^9 \alpha_{0.1}^{-{15\over 32}}\beta_1^{-{3\over
8}}\lambda_\tau^{-{1\over 4}}
m_8^{-{3\over 32}}\[\dot m_{0.1}\phi (1-f)\]^{-1}r_{10}^{75\over 64}\ell_{10}^{1\over 8}K,
\end{equation}  
\begin{equation}\label{e:nr1}
n_1=2.01\times 10^9  \alpha_{0.1}^{-{15\over 16}}\beta_1^{-{3\over 4}}\lambda_\tau^{-{1\over 2}}
m_8^{-{19\over 16}}\[\dot m_{0.1}\phi (1-f)\]^{-2}r_{10}^{75\over 32}\ell_{10}^{-{3\over 4}} {\rm cm}^{-3},
\end{equation} 
where the length of the magnetic loop $\ell_{10}$ is in unit of 
$10R_{\rm S}$.
 
 From Eqs.(\ref{e:energy}), (\ref{e:evap}), (\ref{e:magg}), and (\ref{e:uradre}), we get a coronal solution above a gas
pressure-dominated disk, 
\begin{equation}\label{e:Tg2}
T_2=4.86 \times 10^9  \alpha_{0.1}^{-{9\over 80}}\beta_1^{-{1\over 8}}
\lambda_\tau^{-{1\over 4}}\lambda_u^{-{1\over 4}}
m_8^{{1\over 80}}\[\dot m_{0.1}\phi (1-f)\]^{1\over 10}r_{10}^{-{51\over 160}}\ell_{10}^{1\over 8}K,
\end{equation}
\begin{equation}\label{e:ng2}
n_2=2.55\times 10^{10} \alpha_{0.1}^{-{9\over 40}}\beta_1^{-{1\over 4}}
\lambda_\tau^{-{1\over 2}}\lambda_u^{-{1\over 2}}
m_8^{-{39\over 40}}\[\dot m_{0.1}\phi (1-f)\]^{1\over 5}r_{10}^{-{51\over 80}}\ell_{10}^{-{3\over 4}} {\rm cm}^{-3}.
\end{equation} 
 
The fraction of gravitational energy dissipated
in the corona  can be deduced from its definition (Paper I),
\begin{equation}\label{e:f-d}
f\equiv {F_{\rm cor}\over  F_{\rm tot}}=\({B^2\over 4\pi}V_A\) \({3GM\dot M\phi\over
8\pi R^3}\)^{-1}.
\end{equation}
Replacing  $B$ and $n$ by Eqs.(\ref{e:magr}) and (\ref{e:nr1}), 
we obtain an equation
concerning $f$ for
a radiation pressure-dominated disk,
\begin{equation}\label{e:fr1}
f=1.45(1-f)^{-2}\alpha_{0.1}^{-{45\over 32}}\beta_1^{-{9\over 8}}
\lambda_\tau^{1\over 4}
m_8^{-{9\over 32}}(\dot m_{0.1}\phi)^{{-3}}r_{10}^{{225\over
64}}\ell_{10}^{3\over 8}\equiv c_1(1-f)^{-2}.
\end{equation}
Replacing  $B$ and $n$ by Eqs.(\ref{e:magg}) and (\ref{e:ng2}), we
get an equation  for a gas pressure-dominated disk,
\begin{equation}\label{e:fg}
f=4.70\times 10^4(1-f)^{11\over 10}
\alpha_{0.1}^{-{99\over 80}}\beta_1^{-{11\over 8}}
\lambda_\tau^{{1\over 4}}\lambda_u^{{1\over 4}}
m_8^{{11\over 80}}(\dot m_{0.1}\phi)^{{1\over 10}}r_{10}^{-{81\over
160}}\ell_{10}^{3\over 8}.
\end{equation}  
Eq.(\ref{e:fg}) has solutions for any accretion rate; while
Eq.(\ref{e:fr1}) has solutions only 
if  $ c_1 < 4/27 $, which gives,
\begin{equation}\label{e:mdot-hs}
\frac{\displaystyle \dot M}{\displaystyle \dot M_{\rm Edd}}> 0.21 \alpha_{0.1}^{-{15\over 32}}\beta_1^{-{3\over 8}}
\lambda_\tau^{{1\over 12}}
m_8^{-{3\over 32}}\phi^{-1}r_{10}^{{75\over
64}}\ell_{10}^{1\over 8}. 
\end{equation}
This is to say, at high accretion rate, the corona can coexist
with a radiation pressure-dominated disk or a gas pressure-dominated
disk; while at low accretion rate, the corona can coexist only with the gas
pressure-dominated disk. The value of critical accretion rate is
related to the distances. For $\alpha=0.3$, $\beta=1$, $\lambda_\tau=2.7$ (see next section),
and $M=10^8M_\odot$, a radiation pressure-dominated disk
exists around distance $R\sim 6R_{\rm S}$ if $\dot M\approx 0.3 \dot M_{\rm Edd}$; extends to  
$50R_{\rm S}$ if $\dot M\approx 1.2 \dot M_{\rm Edd}$.
We will see in the following section that
the spectra from a radiation pressure-dominated disk + corona and from
a gas pressure-dominated disk + corona
are of much difference. 
 
Solving  Eq.(\ref{e:fr1}) (or Eq.(\ref{e:fg})) for $f$ for a given black hole mass and
accretion rate, we determine the
coronal quantities from Eqs.(\ref{e:Tr1}) and (\ref{e:nr1}) (or
Eqs.(\ref{e:Tg2}) and (\ref{e:ng2})). The disk quantities can also be
determined. We are now ready to calculate the spectrum.

\section{Spectra calculated from Monte Carlo simulations}\label{s:result}

\subsection{The method of Monte Carlo simulations}
We use the Monte Carlo simulations to 
 calculate the emergent spectrum. Our model consists of a  cold disk, which produces blackbody
radiation at each radius with temperature $T_R$,
\begin{equation}
\sigma T_R^4={3GM\dot M \phi (1-f)\over 8\pi R^3}+{c\over 4}U_{\rm
rad}^{\rm re}\approx \max\left( {3GM\dot M \phi (1-f)\over 8\pi R^3}, {c\over 4}U_{\rm
rad}^{\rm re} \right). 
\end{equation}
The hot corona lying above the disk is approximated  as an
optically thin slab with definite electron temperature and 
scattering optical depth for each radius.
 The blackbody photons emitted from the
cold disk go into the corona, some of them pass through without
being scattered, some of them are Compton up-scattered. By tracing the
photon's motion in the corona, we separately record the photons
emerging from upper and lower boundaries of the slab corona: the
former are the ones to be  observed and the latter impinge the disk
and  are
reprocessed as blackbody emissions. Since the accretion disk is sufficiently
dense, the albedo
is neglected in our simulations. 
 The spectra from different radii are added up as the total emergent spectrum. 
 
The method of the  Monte Carlo simulation is based on Pozdnyakov,
Sobol' \& Sunyaev (1978). We restrict our consideration to a thermal
corona where the electrons have a Maxwellian distribution, and the 
Compton scattering is the main cooling process. Since the
electron-scattering 
optical depth in the corona is less than one, we introduce the weight $w$
as described by Pozdnyakov et al. (1978) in order to efficiently
calculate the effects of multiple scattering. We first set $w_0=1$ for
a given soft photon, then calculate the escape probability $P_0$ of
passing through the slab. The quantity of $w_0P_0$ are the transmitted
portion and is recorded to calculate the penetrated spectrum or
reprocessed photons according to the escape direction of the photon. The
remaining weight $w_1=w_0(1-P_0)$ is the portion that undergoes at
least one scattering. If we write the escape probability after the
$n$-th scattering as $P_n$, the quantity  $w_nP_n$ is the transmitted
portion of photons after the $n$-th scattering, and is recorded as
upward or downward transmitted spectrum. The remaining  portion
$w_n(1-P_n)$ undergoes the $(n+1)$-th scattering. This calculation is
continued until the weight $w$ becomes sufficiently small. The whole
process is simulated by the Monte Carlo method.

\subsection{Iterative computations for consistency}
Once the coronal temperature,  the optical
depth and the seed soft photons are given, the Mento Carlo simulation
can be performed and the Compton scattering
spectrum and luminosity coming out from  both upside 
and downside of the corona are obtained. The questions arise, does
the upward-outcoming luminosity equal to the  gravitation energy
released by the accretion? does the energy density of downward-outgoing photon plus intrinsic disk radiation equal to the
presumed energy density of seed soft photons? If the seed photon energy
and the effective optical depth are correctly chosen in deriving the corona 
temperature,
 density and optical depth, the answers should be
affirmative. A consistent model should correctly describe the nature of corona,
i.e., the temperature, density, optical depth, etc. so that the simulations
give consistent results.

In Paper I, we did not introduce  $\lambda_\tau$  and $\lambda_u$,
which implies  $\lambda_\tau=1$ and
$\lambda_u=1$. This is to say, it is assumed that
the reprocessed seed field has energy density
$U_{\rm rad}^{\rm re}=0.4 {B^2/8\pi}$ and the effective optical
depth is the vertical scattering optical depth
$\tau=n\sigma_T\ell$. However, unlike in a spheric geometry, in a plane
geometry the actual optical depth (we called it the effective optical depth)
should be larger than the vertical one, since the incident photons are
isotropic and undergo longer path than $\ell$ if their directions are
not normal to the  corona plane.  
Furthermore,  $\lambda_\tau$ is expected to be independent on the system
parameters (e.g. accretion rate) since it is only a geometric effect.
The energy density of seed photon
field contributed from
the reprocessed photons, $U_{\rm rad}^{\rm re}=0.4 {B^2/8\pi}$,  is estimated for specific range of parameters (e.g. Haardt \& Maraschi 1993). The precise value depends on
the temperature and optical depth. In order to get  consistent
results, we need to adjust the presumed energy density of soft photons and the
effective optical depth of the scattering medium.  This is why we introduce the
coefficients $\lambda_\tau$  and $\lambda_u$ in this study.   
For simplicity, we set $\lambda_u$ the same value at all distances
though the temperature and density of the corona change a little along
the distance as shown in Paper I. 
This is reasonable since the reprocessed
photons are not required to get out of the disk 
 at the same distance where they impinge.  

For a given mass of black hole and accretion rate, we first perform
the computations for the case of gas pressure-dominated 
disk. As the first step, we start the
computation with initial values  $\lambda_\tau=2$ and
$\lambda_u=1$. For one grid point of 
radius $R$,  we solve 
Eq.(\ref{e:fg}) numerically and obtain a solution for  $f$; then  calculate the corresponding temperature $T$ and density
in the corona from
Eqs.(\ref{e:Tg2}) and (\ref{e:ng2}) and hence the scattering optical depth $\tau$, and the radiation temperature $T_R$ of seed photon field by combining
Eqs.(\ref{e:uradre}) and (\ref{e:magg}). With $T$, $\tau$, and $T_R$,
we perform the Monte Carlo simulation, record the local upwards
luminosity and downwards luminosity. Then we change $R$ to the next grid point
and repeat the calculation for $f$, then for  $T$, $\tau$ and
$T_R$, and then Monte Carlo simulation. Finally
all the local outgoing photons are added up and the integrated upward
luminosity $L_{\rm up}$ and downward luminosity $L_{\rm down}$ are
obtained. Meanwhile, the integrated soft photon energy $L_{\rm
soft}=\int_{3R_{\rm S}}^{50R_{\rm S}}2\pi R\sigma T_R^4dR$ and the released
gravitational energy $L_{\rm G}=\int_{3R_{\rm S}}^{50R_{\rm
S}}2\pi R(3GM\dot M\phi/8\pi R^3)dR$ are also calculated.
The second step is  to check the consistency, i.e. 
 $L_{\rm down}\approx L_{\rm soft}$ and $L_{\rm
up}\approx L_{\rm G}$? If yes, we find the correct  $\lambda_\tau$  and
$\lambda_u$ and hence the consistent solution for the corona, and stop the computation; if no, set new values  for
 the $(n+1)$-th $\lambda_\tau$  and $\lambda_u$ from the $n$-th results:
$\lambda_{u,n+1}=(L_{{\rm down},n}/L_{{\rm soft},n})\lambda_{u,n}$ and 
$\lambda_{\tau,n+1}=(L_{{\rm up},n}/L_{\rm
G}) \lambda_{\tau,n}$, then repeat the computation from the first step until
the consistent conditions $L_{\rm down}\approx L_{\rm soft}$ and $L_{\rm
up}\approx L_{\rm G}$ are fulfilled.

In the case of radiation pressure-dominated disk,  We  solve $f$ from
Eq.(\ref{e:fr1}) and then  the temperature and density from
Eqs.(\ref{e:Tr1}) and  Eqs.(\ref{e:nr1}).
The iteration procedure is similar to the case of gas pressure-dominated
disk.
The difference is in that we only need to check whether  $L_{\rm up}\approx L_{\rm
G}$ is fulfilled or not,  since the contribution to
the seed photons from the reprocessed
photons are not so important as the intrinsic disk
radiations. Thus,  $\lambda_u$ is simply set to unity and is not
adjusted during the iteration. 

\subsection{The self-consistent coronal structure}\label{s:structure}
\begin{figure}
\plotone{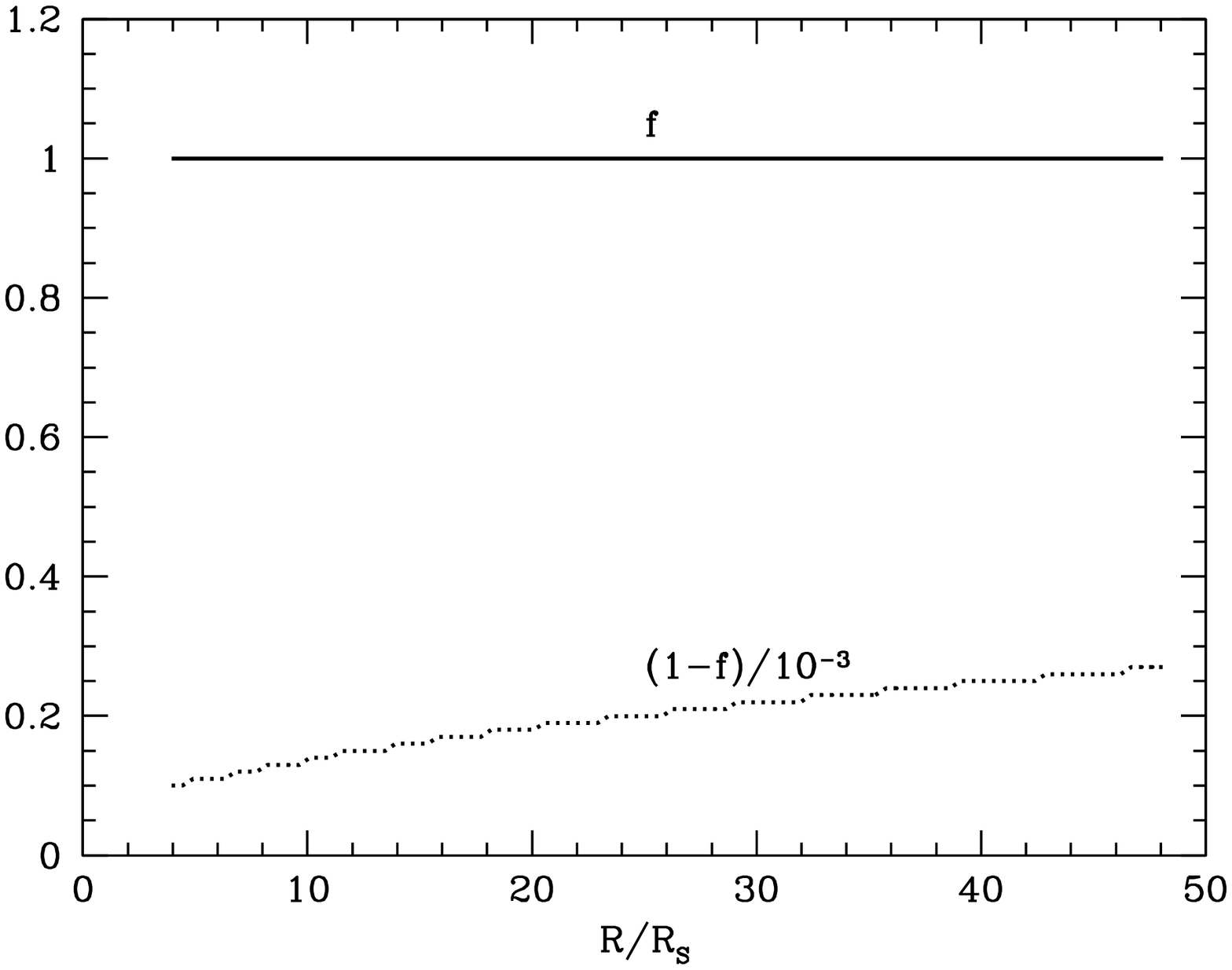}
\caption[]{\label{f:fmdot01} Accretion-energy fractions
dissipated in the corona, $f$, and in the disk, $1-f$, for
 $M=10^8M_\odot$ and $\dot M=0.1\dot M_{\rm Edd}$ (corresponding to
 $L=0.07L_{\rm Edd}$).  The disk is
 gas pressure-dominated.
The figure shows that most of the accretion energy is dissipated in the
 corona. Only a very small fraction is dissipated in the disk.}
\end{figure}

For given parameters, $M=10^8M_\odot$, $\alpha=0.3$,
$\ell=10R_{\rm S}$, $\beta=1$, we perform computations for a series of 
accretion rates. As we discussed in Sect.\ref{s:model}, computations
show that,  for low accretion rates ($\dot M\la 0.3
\dot M_{\rm Edd}$) there is only
gas pressure-dominated solution;  for high
mass accretion rates there are both gas
pressure-dominated and radiation pressure-dominated solutions 
(say, $\dot M=2\dot M_{\rm Edd}$); for 
middle  accretion rates (say, $\dot M=0.5\dot M_{\rm Edd}$), the gas pressure-dominated solution exists
at all distances, while the
radiation pressure-dominated solution only exists at small distances.  As an example for low accretion rate system, Fig.\ref{f:fmdot01} shows how much accretion energy is dissipated in the
corona and in the disk for 
$\dot M=0.1\dot M_{\rm Edd}$. It is shown that almost all the accretion
energy is dissipated in the corona; less than one thousandth is
dissipated in the disk.  The corresponding coronal structure is
plotted in
Fig.\ref{f:tnyhard}.
 The temperature is $\sim 10^9$K and the density $\sim 10^9$cm$^{-3}$, which are
a little lower than that in Paper I since here we take into account the
effective optical depth and consistent seed photon field. The effect
optical depth is around 0.7 and Compton y-parameter 0.6.

\begin{figure}
\plotone{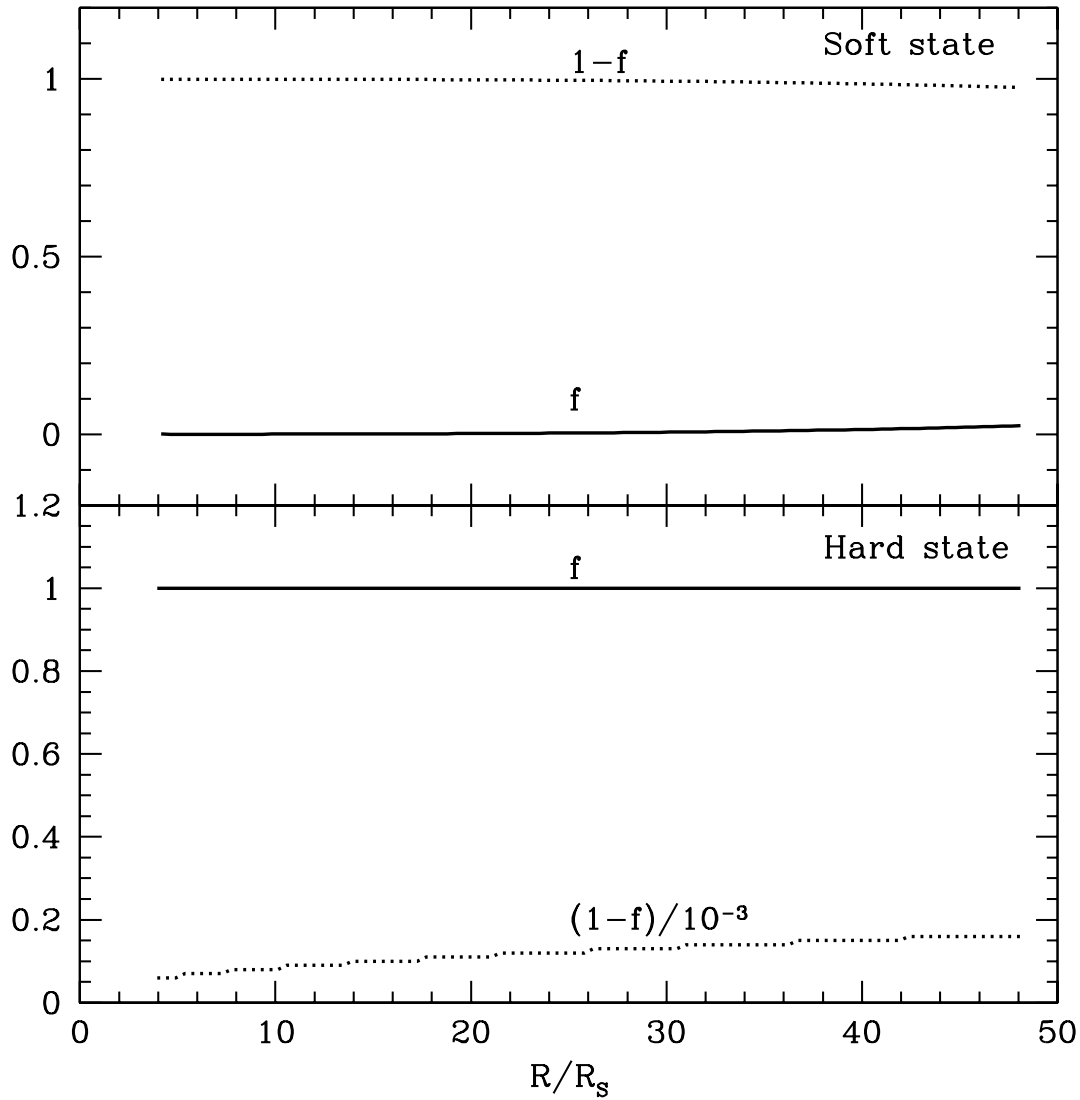}
\caption[]{\label{f:fmdot2} Two solutions of $f$  for
 $M=10^8M_\odot$ and $\dot M=2.0\dot M_{\rm Edd}$ 
(corresponding to $L=1.4L_{\rm Edd}$). 
The lower panel shows that for a gas pressure-doimnated disk (hard state)
 a large fraction of accretion energy is
 dissipated in the corona;
 The upper panel shows that for a radiation pressure-doimnated disk 
(soft state) only a
 very small fraction of accretion energy is transported to the corona,
 while most of the energy is dissipated in the disk.}
\end{figure}

\begin{figure}
\plotone{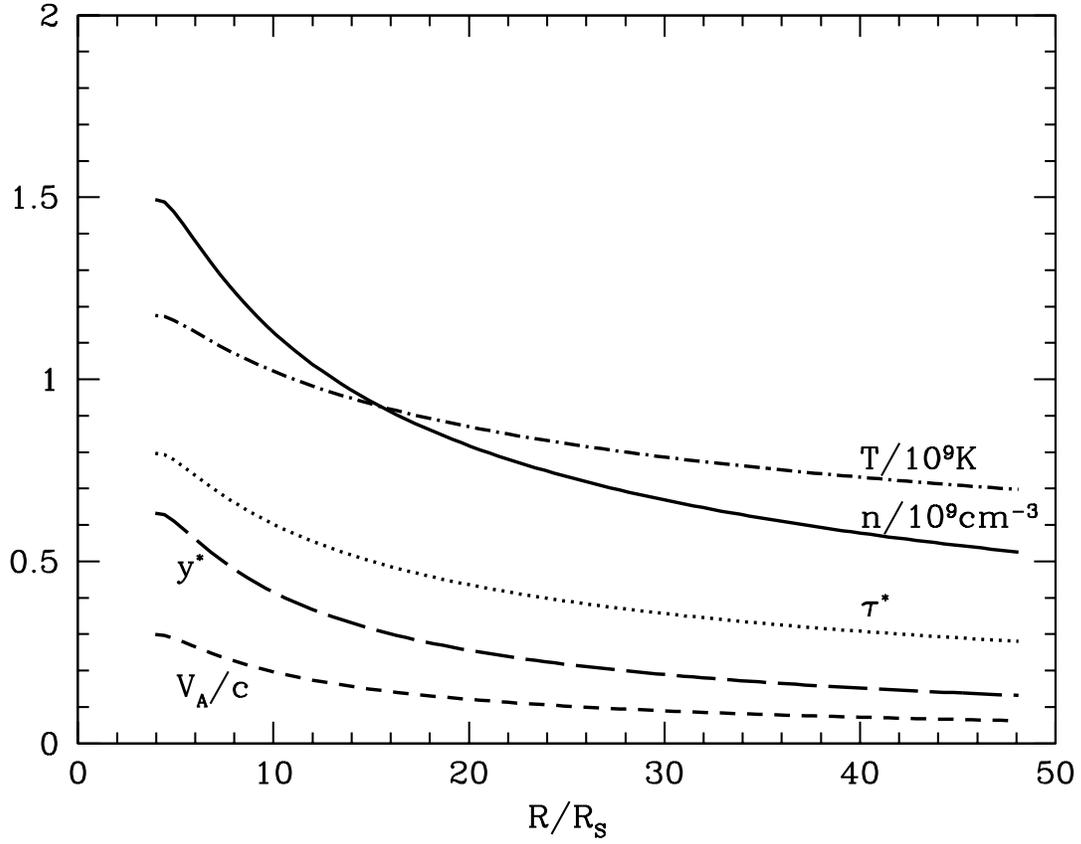}
\caption[]{\label{f:tnyhard} Coronal quantities along distance for
 $M=10^8M_\odot$ and $\dot M=0.1\dot M_{\rm Edd}$ (corresponding to
 $L=0.07L_{\rm Edd}$).
The corona is above a gas pressure-dominated disk with energy fraction
channeled to the corona $f\sim 1$ shown in Figure \ref{f:fmdot01}.
The coronal temperature is around $10^9$K and density
around $10^9$cm$^{-3}$.  The effective optical depth $\tau^*$ and
Compton $y$-parameter $y^*$ are also shown,
which hardly change with the luminosity.}
\end{figure}

\begin{figure}
\plotone{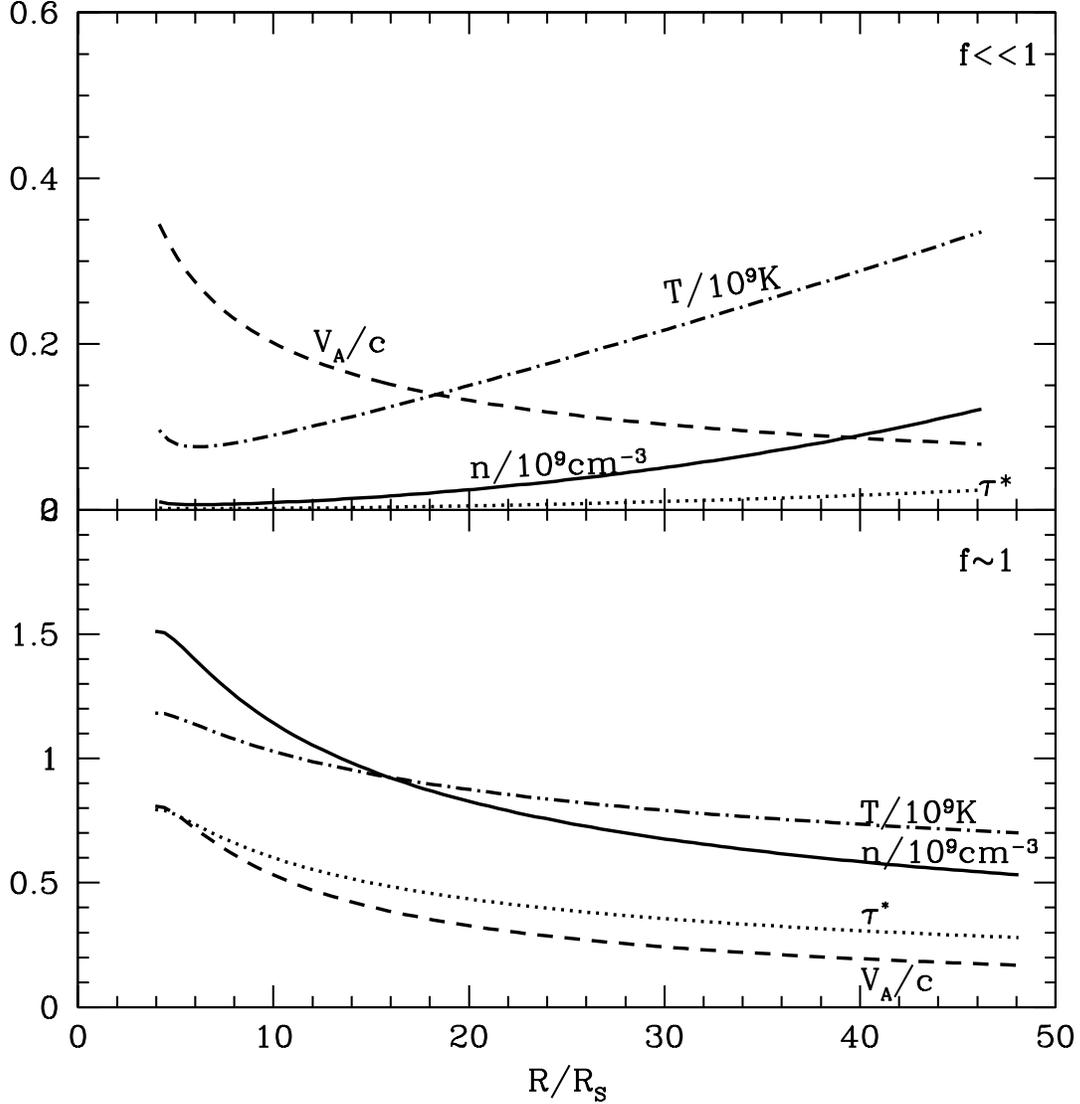}
\caption[]{\label{f:tnysoft} Coronal quantities along distance for
 $M=10^8M_\odot$ and $\dot M=2.0\dot M_{\rm Edd}$ 
(corresponding to $L=1.4L_{\rm Edd}$). 
The lower panel shows structure of the corona above a gas pressure-dominated disk with energy fraction
channeled to the corona $f\sim 1$. The temperature and density are
almost the same as that for small $\dot M$ shown in Figure \ref{f:tnyhard};
while the Alfv\'en speed is
larger at higher $\dot M$.
The upper panel shows structure of the corona above a radiation pressure-dominated disk.
Since most of the accretion energy is dissipated in the disk,  the
corona is weak with temperature $\sim$ a few $10^8K$ and density
$\la 10^8 {\rm cm}^{-3}$ in the inner region. In the outer
region,  $T$ and $n$ are relatively high due to a little larger
fraction of local accretion energy is transfered to the corona.}
\end{figure}

As an example for
high accretion rate, Fig.\ref{f:fmdot2} and Fig.\ref{f:tnysoft} show the energy fraction and coronal structure for an
accretion rate of  $2\dot M_{\rm Edd}$. For a gas pressure-dominated disk, the coronal
temperature and density are almost the same as that for low accretion
rates. Only the magnetic field and Alfv\'en speed increase with
accretion rates. This feature is the nature of gas-pressure
solutions because $f$ depends on $\dot M$
very weakly (see Eq.(\ref{e:fg})), so are 
the temperature (see Eq.(\ref{e:Tg2})) and  density (see Eq.(\ref{e:ng2})).    
In other words, the coronal structures with a gas pressure-dominated disk 
underneath are quite uniform. Higher
accretion rate only results in faster energy transfer and dissipation
in the corona. Such  features can be understood as follows.
A corona  coupled with the disk through the magnetic field and
radiation field can self-adjust to an equilibrium at a
certain temperature. The magnetic field transports almost all the
accretion energy to the corona and releases the energy there through
reconnection.   At the beginning of reconnection, the electrons are
suddenly heated to a very high temperature. Even though there is no many seed
photons to be scattered, the Compton cooling power is still quite
strong due to the very high temperature. Thus, the seed photon's energy  
is greatly amplified. Some fraction of the scattered photons get
out of the corona from the backside and are reprocessed in the
disk   as blackbody
radiation. Then the seed photon field is strengthened and the electron
temperature decreases due to efficient Compton cooling. The equilibrium is
finally reached when the heating by the magnetic reconnection
balances  the Compton cooling. 
Such an  approach to the equilibrium can be clearly seen from
our iterative computations: when $\lambda_u$ is set to  be a small value,
the temperature derived from the model is high. With this high $T$ we
then obtain high backward-scattered photon luminosity (by the
Monte Carlo simulation). Taking this high luminosity
as the seed soft photon luminosity, the newly derived $T$ becomes low and
hence the soft-photon field contributed from the backward Compton radiation
is closer to the true value. 
By repeating this procedure, the
backward-scattered luminosity quickly converges to the seed soft
luminosity, and a stationary state is reached.

In table \ref{t:lambda} we list for a gas pressure-dominated disk 
the coefficients $\lambda_u$,
$\lambda_\tau$, the emergent luminosity $L_{\rm up}$, the seed soft photon
luminosity $L_{\rm soft}$,  and the amplification factor $A\equiv
(L_{\rm up}+L_{\rm down})/L_{\rm soft}$, where $L_{\rm down}$ is the
downward/backward Compton luminosity. 
We find that the Compton amplification factors for
different accretion rate are almost the same, $A\sim 2.3$.  
Soft photons are scattered in the corona, 
gain energy by a factor of $A-1\sim 1.3$ and escape from both sides of the corona. The
upward-escaped photons carry a little  more energy ($L_{\rm up}$) than the
downward-escaped photons ($\approx L_{\rm soft}$), the latter is
reprocessed as seed photons.
The coefficient $\lambda_u$ for the seed photon
field increases with increasing accretion rate.  Comparing $\lambda_u$
and Alfv\'en speed, we find that $\lambda_u/V_A$ does not change with
$\dot M$. This indicates that the reprocessed photon field is actually proportional
to the $U_{\rm B}(V_A/c)$, the factor $V_A/c$ is due to the lag of magnetic
heating (at a speed of $V_A$) to radiation cooling (at a speed of $c$).
We also find $\lambda_\tau\approx 2.7$,  independent on  $\dot
M$. This is to say, the effective optical depth caused by the slab geometry
is as 2.7 times large as the vertical optical depth.  

\begin{table}[h]
\begin{center}
\caption[]{\label{t:lambda} Computational results
 for a corona above a gas pressure-dominated disk at different
accretion rates (where $\dot M_{\rm Edd}\equiv L_{\rm Edd}/0.1c^2$)}
\begin{tabular}{cccccc}
\noalign{\smallskip}\hline\hline\noalign{\smallskip}
$\dot M/\dot M_{\rm Edd}$&$A$&$\lambda_u$&$\lambda_\tau$ &$L_{\rm up}$  & $L_{\rm soft}$\\
0.01 &2.35&1.10&2.74& $4.43\times 10^{43}$& $3.30\times
10^{43}$\\
0.05 &2.34&1.89&2.72&$2.21\times 10^{44}$& $1.65\times 10^{44}$\\
0.1 &2.34&2.38&2.71&$4.43\times 10^{44}$& $3.30\times 10^{44}$\\
0.5&2.33&4.08&2.68&$2.22\times 10^{45}$& $1.66\times 10^{45}$\\
1.0&2.34&5.12&2.68&$4.45\times 10^{45}$& $3.30\times 10^{45}$\\
2.0&2.35&6.40&2.68&$8.84\times 10^{45}$& $6.57\times 10^{45}$\\
\noalign{\smallskip}\hline\hline\noalign{\smallskip} 
\end{tabular}
\end{center}
\end{table}

In contrast, for a radiation pressure-dominated disk, 
  a large fraction of the accretion energy
dissipates in the disk (see Fig.\ref{f:fmdot2}), the
corona is quite weak with temperature being a few $10^8K$ and density 
$\sim 10^8 {\rm cm}^{-3}$ (see Fig.\ref{f:tnysoft}). 
 The seed photons are mainly from the intrinsic
disk radiation. The disk and corona are not coupled at the same way as
that for a gas pressure-dominated disk.
In such a strong disk + weak corona system,
the Compton scattering in the corona is quite
weak, $\lambda_\tau \approx 1$ and $A\approx 1$.
We note that, the energy fraction dissipated in the corona 
decreases with increasing accretion rate, the temperature and density
in the corona also decrease with $\dot M$. This implies that
the corona above a radiation pressure-dominated disk
is weak at luminous systems.

\subsection{The emergent spectrum }\label{s:spectrum}   
\begin{figure}
\plotone{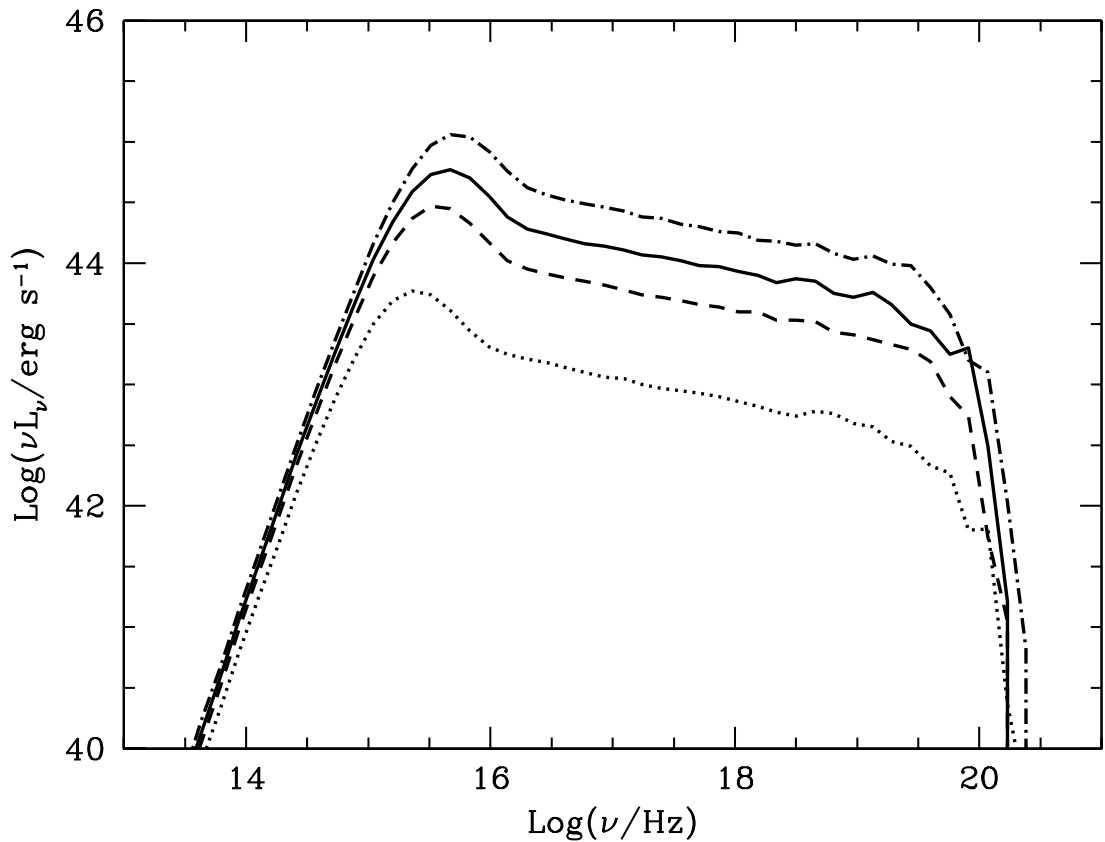}
\caption[]{\label{f:spect-hard}The hard-state spectra from a disk and corona
around a black hole of $10^8M_\odot$ at accretion rates
 $\dot M=0.1\dot M_{\rm Edd}$ (or
 $L=0.07L_{\rm Edd}$; Dotted curve);  $\dot M=0.5\dot M_{\rm Edd}$  (or $L=0.35L_{\rm Edd}$; Dashed curve);
 $\dot M=1.0\dot M_{\rm Edd}$ (or $L=0.7L_{\rm Edd}$; Solid curve); and 
$\dot M=2.0\dot M_{\rm Edd}$ (or $L=1.4L_{\rm Edd}$; Dash-dotted curve). 
The hard X-ray emission is quite
 strong with a spectral index around 1.1. The spectral energy
 distributions are similar for different luminosities.}
\end{figure}

In last section we show that there are two types of solution for the disk and corona structure. 
When the accretion disk is presumed to be gas pressure-dominated, we
get a solution $f\sim 1$, indicating that 
 most of the gravitational
energy is transferred to the corona by strong magnetic field and fast
reconnection. The disk is quite cool, with midplane temperature
of a few
$10^4$K. The corona is heated up to a temperature $T\sim
10^9$K by the magnetic reconnection, and the density
becomes $10^9$cm$^{-3}$ through efficient mass evaporation 
at the interface between the disk and the corona.
The dissipated energy in the corona is emitted away by Compton
scattering. Part of the seed photons 
are upward scattered as the emergent spectrum, 
and part of them
are scattered backward, which are reprocessed in
the disk surface layer and are emitted as blackbody spectra.
 Here we assume that
the irradiation by the backward Compton radiation does not largely influence
the internal structure of a radiative disk  but only changes the
temperature of the surface layer (for details see Tuchman,
Mineshige, \& Wheeler 1990). Therefore, the magnetic energy determined by the
equipartition  in the disk midplane is not affected by the irradiation. 
Such a highly heated corona produces
strong hard X-ray emission. The corona and disk are similar to the
``two phases'' discussed by Haardt and Maraschi (1993).
We call this solution as hard state since its spectrum is hard. 
Fig.\ref{f:spect-hard} shows the spectral energy distribution of 
hard state for different accretion rates corresponding to different luminosities. [Here we need to
point out that the energy conversion coefficient $\eta$ by accretion is not exactly 0.1 in our case. For
given accretion rate we integrate the gravitational energy by $L_{\rm
G}=\int_{R^*}^{50R_{\rm S}}
2\pi R(3GM\dot M\phi/8\pi R^3)dR$. The total luminosity from both side of the disk and corona 
should be $2L_{\rm G}$.  For $R^*=3R_{\rm S}$,
$\eta=2L_{\rm G}/\dot M c^2\sim 0.07$. Thus, even for $\dot M=\dot M_{\rm Edd}\equiv L_{\rm
Edd}/0.1c^2$, the luminosity is less than the Eddington luminosity.]
From the figure we see that for different $\dot M$ the spectral shapes in the hard state
are very similar. The hard X-ray spectral index is around 1.1, which
 is a little
steeper than that observed in Seyfert galaxies. 

The ``constant'' spectral index is caused by  the ``uniform'' corona. As we
show in Sect.\ref{s:structure}, the coronal structure at hard state 
 is nearly independent on
the accretion rate and the Compton $y$-parameter ($\equiv (4kT/m_ec^2)\tau$) hardly changes
with  $\dot M$.
 Our computational data show that the
radial distributions of  $y$-parameter are indeed the same for different  $\dot
M$; $y$ reaches the maximum of  $y\sim 0.6$ in the inner region and
decreases with distances (see Fig.\ref{f:tnyhard}). Therefore, the hard-state spectra have 
similar spectral shapes.  

\begin{figure}
\plotone{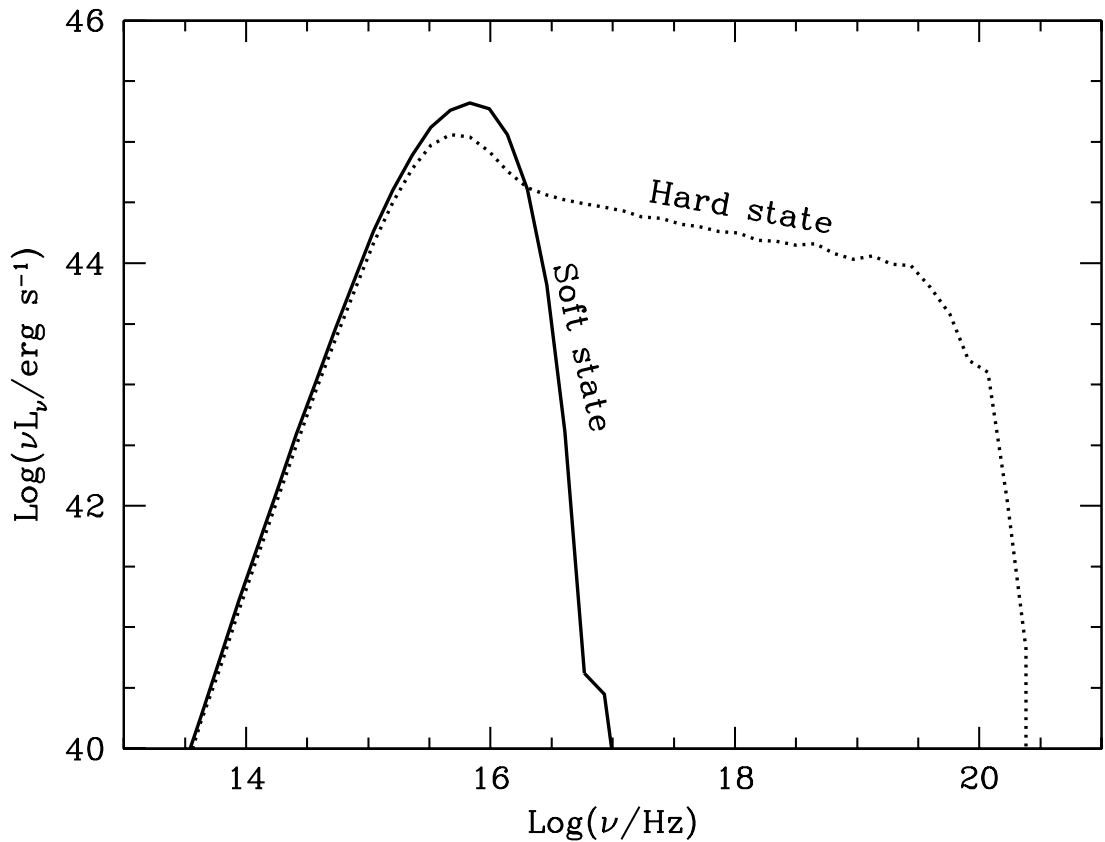}
\caption[]{\label{f:spect-soft}The soft-state spectrum (Solid curve) from a disk and corona
around a black hole of $10^8M_\odot$ at accretion rate $\dot M=2.0\dot M_{\rm Edd}$ (or $L=1.4L_{\rm Edd}$). The spectral energy
distributes dominantly in UV and soft X-ray
since most of the accretion energy is dissipated in the disk. For
comparison the
corresponding hard-state spectrum is also shown in the figure (Dotted curve).
Obviously, the X-ray spectra at soft state and hard state are of much difference.}
\end{figure}

On the other hand, if the disk is presumed to 
be radiation pressure-dominated, there exists no solution for $f$ at
low accretion rate. At high accretion rate, the solution for $f$ is
far less than 1, indicating that   
magnetic buoyancy does not transfer much energy from the disk to
the corona, and there is only a weak corona above the disk with low
temperature and low
density  compared to those in the hard state. 
 The gravitational
energy is dominantly released in the disk as  multi-color blackbody
radiation. These photons go through the very weak corona, almost all of
them penetrate the corona without being
scattered. The spectral energy ($\nu L_\nu$) peaks at around UV to soft
X-rays, dropping steeply at high frequency.
 We call this solution as the soft state. Fig.\ref{f:spect-soft} shows 
typical 
spectrum of the soft state for $\dot M=2\dot M_{\rm Edd}$ (corresponding to
$L=1.4L_{\rm Edd}$). The soft-state spectrum is not much different from 
pure-disk spectrum.
For comparison the hard-state spectrum for the same $\dot M$ is also shown in the
figure. Obviously, the spectral distributions at soft and hard state 
are very much different.
 One may question whether the Bremsstrahlung emissions contribute more to the
hard X-rays in this state. Our estimation shows that the Bremsstrahlung emission power
is about 5 order of magnitude less than Compton radiation because the density in the
soft-state corona is so low.   

\begin{figure}
\plotone{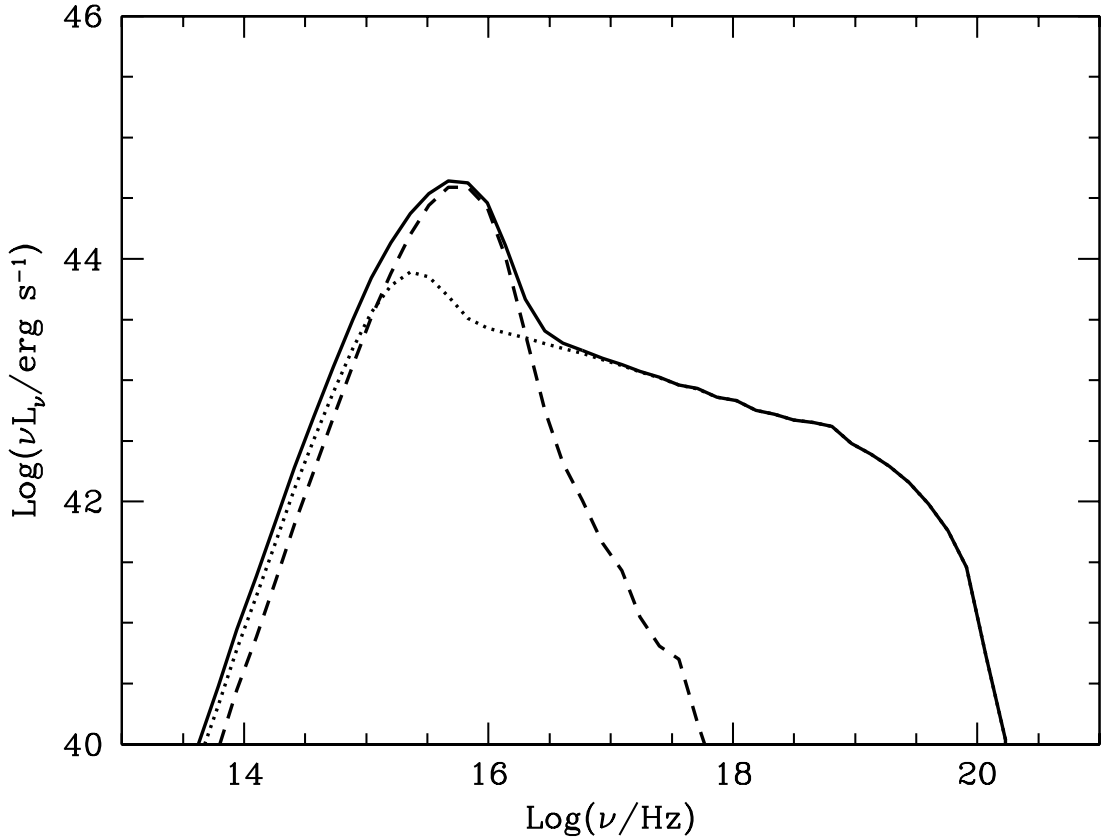}
\caption[]{\label{f:spect-hs} The composed  spectrum from a disk and corona
around a black hole of $10^8M_\odot$ at accretion rate $\dot M=0.5\dot
M_{\rm Edd}$  (or $L=0.35L_{\rm Edd}$). The disk is radiation 
pressure-dominated in the inner region until $19R_{\rm S}$ and gas
pressure-dominated in the outer region, and hence the overlying corona 
is weak in the inner region and strong in the outer region.
Dashed curve shows spectral energy distribution from the inner region, 
which is obviously contributed by the disk radiation; 
Dotted curve shows the contribution by the outer region, which
is a typical radiation-coupled hard-state spectrum;
The solid curve shows the total spectrum. 
The overall spectral energy distribution is between the
hard sate and soft state, and the softness increases with accretion rate.}
\end{figure}

Furthermore, there are composite solutions, in which the disk is composed of an inner radiation
pressure-dominated region and an outer gas pressure-dominated region
with a weak inner corona and a strong outer corona above. Whether such a
discontinuous disk + corona can dynamically exist is to be
studied. Nevertheless, the spectra from such a composite corona + disk
show moderate UV-X-ray spectral indices.  
In Fig.\ref{f:spect-hs} we plot  such a spectrum for $\dot
M=0.5\dot M_{\rm Edd}$
(corresponding to $L=0.35L_{\rm Edd}$).  It is shown that the inner region
produces soft spectra dominated by the disk radiation and
the outer region produces typical
hard-state spectra dominated by the corona radiation. Thus, 
the composed spectrum is moderate: its hard X-ray spectrum has the same 
spectral index as that of hard state; while the ratio of UV and X-ray 
luminosity, or rather, the UV-X-ray spectral index, is larger than that 
of hard state. 
With $\dot M$ increases, the inner radiation pressure-dominated region
extends outwards and the outer gas-pressure dominated region
shrinks. The overall spectrum then becomes softer.
 
As discussed in Sect.\ref{s:structure},  the hard-state solution can exist in both low- and high-luminosity
 objects; while a soft-state corona extending to 50$R_{\rm
 S}$ can only appear at high accretion rate,
 $\dot M\ga 1.2\dot M_{\rm Edd}$ (or $L\ga 0.8L_{\rm Edd}$).  Systems accreting at a rate
$0.3\dot M_{\rm
 Edd} \la \dot M \la 1.2\dot M_{\rm Edd}$ (or $0.2L_{\rm Edd}\la L\la 0.8L_{\rm Edd}$)
can  be in  a state between hard and soft with UV-X-ray indices varying with
 accretion rates.
Therefore, our model predicts that, two spectral states are possible
 for accretion rate above $1.2 \dot M_{\rm Edd}$; below  $0.3 \dot M_{\rm Edd}$   only hard
 state.
These features of hard and soft states are summarized in Table \ref{t:state}.
\begin{table} [h]
\begin{center}
\caption[]{\label{t:state} The features of disk and corona in the hard and soft spectral states}
\begin{tabular}{ccccccccc}
\noalign{\smallskip}\hline\hline\noalign{\smallskip}
Spectral&Range  &$f$& Pressure&\multicolumn{3}{c}{Typical coronal values}&$\alpha_{\rm px}$& $\alpha_X$\\
state& of $L$&&in disk& T(K)& n(cm$^{-3}$)&$\tau^*$&\\
\noalign{\smallskip}\hline\noalign{\smallskip}
Hard&Always&$\sim 1$&$P_{\rm disk}^{\rm g}$&$\sim 10^9$&$\sim 10^9$&0.6&$\sim 1.4$&$\sim 1.1$\\
Soft& $\ga 0.8L_{\rm Edd}$&$<0.1$&$P_{\rm disk}^{\rm
r}$&\multicolumn{3}{c}{Low, decrease with $L$}&$> 2$&\\
Moderate&0.2-0.8$L_{\rm Edd}$&\multicolumn{5}{c}{outer region hard,
inner region soft}&$>1.4$&$\sim 1.1$\\
\noalign{\smallskip}\hline\hline\noalign{\smallskip} 
\end{tabular}
\end{center}
\end{table}

\begin{figure}
\plotone{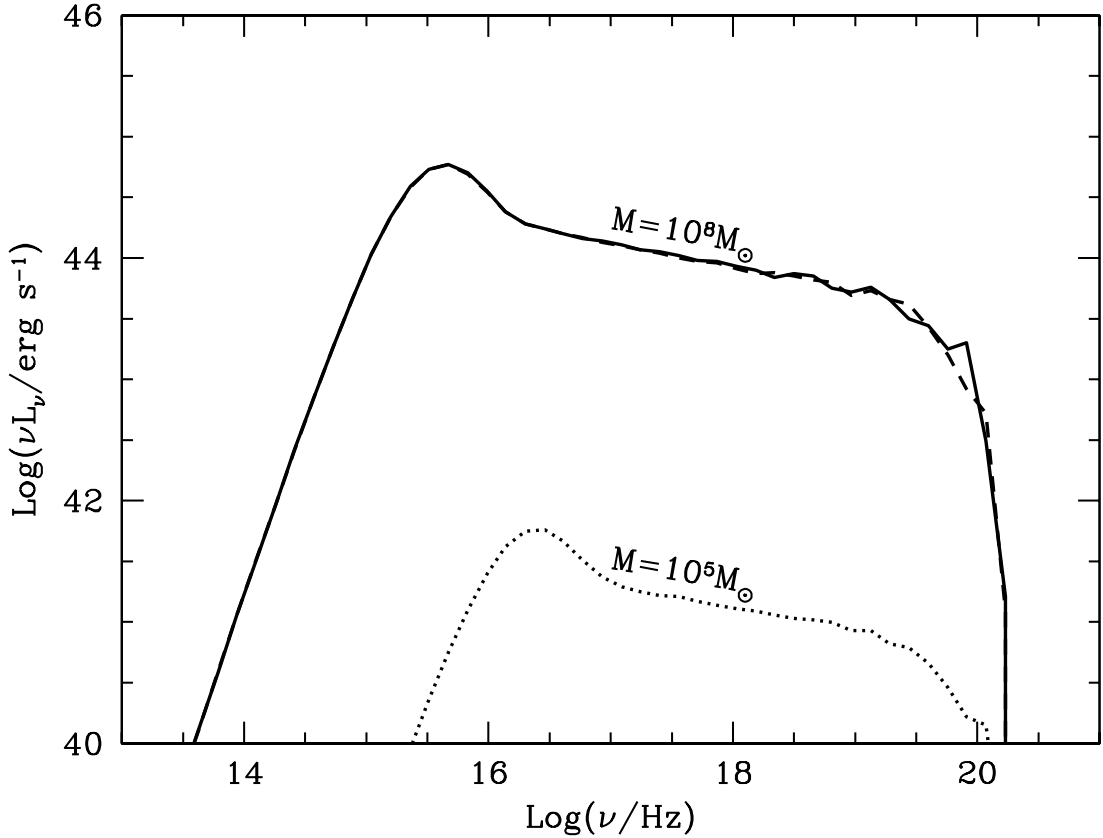}
\caption[]{\label{f:spect-ml}The hard-state spectrum for
$M=10^5M_\odot$ and $\dot M=1\dot M_{\rm Edd}$ (or $L=0.7L_{\rm
 Edd}$) (Dotted curve) compared to that for 
 $ M=10^8M_\odot$  at same accretion
 rate $\dot M=1\dot M_{\rm Edd}$ (Solid curve). 
The spectral shapes are similar except that the disk radiation for
 $M=10^5M_\odot$ is at higher frequency  due to higher disk 
 temperature for lower black-hole mass.  
 An example of $\ell=20R_{\rm S}$ for  $ M=10^8M_\odot$ and $\dot
 M=1\dot M_{\rm Edd}$ is also shown in the figure (Dashed curve), which
  overlaps in the standard case ($\ell=10R_{\rm S}$). This indicates
 that  the spectrum hardly changes with the thickness of corona since the
  hard-state temperature and optical depth  given by our model depend on $\ell$ very weakly.}
\end{figure}

The parameters in our model are the mass of black hole $M$, the accretion
rate $\dot M$, the thickness of corona $\ell$ (equivalent of averaged length of magnetic loops), and the equipartition
coefficient $\beta$. Above study shows  the
corona structure and spectra in dependence on $\dot M$ for given parameters
$M=10^8 M_\odot$ and $\ell=10R_{\rm S}$ in the region from 3$R_{\rm
S}$ to 50$R_{\rm S}$. From the equation concerning $f$ and the
solutions of T and $\tau (\propto n\ell)$, we know $f$, $T$ and $\tau$ and hence the spectra only very weakly depend on $\ell$,
especially at the hard state. In Fig.\ref{f:spect-ml} we plot the
hard-state spectra for $\ell=10R_{\rm S}$ and $\ell=20R_{\rm S}$. The
two curves overlap. Similarly, the spectra have weak dependence on the
mass of black hole. But, considering large change of black-hole mass over a few orders of magnitude,
the spectra may be somewhat different. In
Fig.\ref{f:spect-ml} an example of hard state for $M=10^5 M_\odot$ is shown. 
 Compared to the case for $M=10^8 M_\odot$, the disk component  shifts towards high frequency, 
whereas no obvious difference in the shape of hard X-ray spectra contributed by the corona. 
For a $10M_\odot$ black hole we expect
 similar hard X-ray spectra except for the
low-frequency component which peaks at soft X-rays due to high effective temperature in the disk. The dependence on $\beta$
seems large because it directly determines  how much of the accretion
energy transferred to the corona through magnetic field. Owing to
poor knowledge on the equipartition coefficient, we don't discuss this
issue here.       

\section{Discussion on the spectra}\label{s:discussion}
The observed spectrum in X-ray 2-20 keV is close to a power
law, with an averaged spectral index $\approx 0.7\pm 0.15$  for Seyfert
galaxies
(Mushotzky 1984; Turner \& Pounds 1989). Taking into account
the deconvolution of the Compton reflection hump around $\approx$30keV
(George \& Fabian 1991; Williams et al. 1992; Mushotzski et al. 1993; 
 Petrocci et al. 2000), 
the spectral index of the
underlying power law for Seyfert galaxies is 0.9 to 1.0 
(e.g. Nandra \& Pounds 1994; Pounds et al. 1990). 
The X-ray spectral index for radio-quiet QSOs is slightly higher, 
$\alpha_X \approx 1.0$ (e.g. George et al. 2000). 
The observed broad band spectral index  measured between
2500$\AA$ and 2keV is 1.25 for Seyferts (Walter \& Fink 1993)
 and 1.5 for QSOs (Yuan et al. 1998). 
Our model shows that at hard state the X-ray spectral indices between 2-20keV
are around 1.1. The spectral indices between the hump (around UV) and 2keV,
$\alpha_{\rm px}$, are around 1.4. The hard-state spectra are roughly 
in agreement
 with  the observed spectra of Seyfert galaxies and QSOs.

The composite spectrum, which occurs only at relative high accretion rates
 and becomes more and more close to the soft spectrum with increase
 of $\dot M$,   may explain the diversity of spectra observed in 
narrow-line Seyfert 1 galaxies (NLS1s). NLS1 is a
subgroup  of Seyfert 1 galaxies, but is characterized by large soft
X-ray excess (see Fig.1 in Boller 2000). It has been proposed that
black hole masses are systematically smaller, while luminosities are
comparable to those of other Seyfert 1s, leading to large 
$L/L_{\rm Edd}$  in NLS1s (Osterbrock \& Pogge 1985; Boller, Brandt,
\& Fink 1996; Mineshige et al. 2000). Our composite soft spectra may
correspond for various spectra observed in NLS1s.  
Furthermore, a small fraction
of QSOs with optical-X-ray spectral index $\alpha_{\rm oX} \ga 2$  are
 observed (Yuan et al. 1998), 
which may  be interpreted by our composite  soft spectra.
Comparison with GBHCs are another interesting, outstanding issue but is
beyond the present investigation.

Compared with the excellent work of Haardt \& Maraschi (1993,
hereafter HM93), our
hard-state results are quite similar to theirs. This is easy to understand,
 since our study on the interaction between the disk and corona shows $f\sim 1$, 
which is a basic assumption in HM93. We also find soft-state spectra, 
where the corona is not coupled with the disk in the same way as that in HM93. 
Therefore, the spectra are different from HM93.   
 Despite of the
complexity, the basic differences can be briefly described as follows. HM93 consider the
coupling between the disk and the corona through seed photons, which gives a
relation between the soft radiation temperature $T_R$, the corona
temperature $T$, and the optical depth $\tau$, i.e. $F_1(T_R, T,
\tau)=C_1$. They find for $5eV<T_R<50eV$ and $0.01<\tau<1$, the
relation can be approximated as $(16\Theta^2+4\Theta)\tau=0.6$
(where $\Theta\equiv kT/m_ec^2$). For given $T_R$ and $\tau$, $T$ is
determined by this approximation, and then the spectrum is obtained
from the Monte Carlo simulations.  By changing $T_R$ and $\tau$ within
the limited range, they got somewhat similar spectral indices.  In our
model, we also consider the energy coupling between the disk and
corona through the soft photons, i.e. $F_1(T_R, T, \tau)=C_1$. In
addition, we consider the coupling by the magnetic field and get the
second relation, $F_2(T_R, T, \tau)=C_2(M, \dot M, f)$. Supplemented by
the mass evaporation, we have the third relation, $F_3(T,
\tau)=C_3$. Therefore, we uniquely determine $T_R, T,$ and $ \tau$ for
given $f$, $M$, and $\dot M$. Then, we perform the Monte Carlo simulation and obtain the
spectrum and upward luminosity $L(M,\dot M,f)$. Using the fourth relation
of that the luminosity equals to the accretion energy, i.e., $L(M,\dot M,f)=L_G(M,\dot M)$, 
we can also determine $f$. HM93 also take into account
other factors like anisotropy of the seed photons. Here we concentrate
on the underlying physics in order to determine the corona variables.     

\section{Conclusion}\label{s:conclusion}

We improved our simple model presented in Paper I by introducing
adjustable coefficients for the energy density of the 
coupled seed photons and  
for the optical depth of plane-parallel corona. For given mass of black
hole, accretion rate, and presumed coefficients, the model can
determine the energy fraction dissipated in the corona and hence
determine the structure of both the corona and disk.
 Then, we perform
 Monte Carlo simulations to trace the motion of photons in the corona and record the
upward photons (emergent luminosity) and the backward photons. By
comparing the presumed seed photon energy and the sum of
intrinsic disk radiation and backward radiation energy, we adjust the
 coefficient of soft photon field and; by comparing the total accretion
energy and the simulating emergent luminosity, we modify the
effective optical depth in the slab geometry. Repeating 
this procedure, we finally obtain the
self-consistent solutions for the corona and also for the disk, both
of which
are coupled by the magnetic field and the radiation field.

We find two types of solutions for the disk and corona structure,
 corresponding for the hard-state and
 soft-state spectra. In the hard-state solution, the accretion disk is 
 gas pressure-dominated. Most of the accretion energy is transferred
to the corona by the magnetic field, and the intrinsic disk radiation is very
weak.  The corona and disk are tightly coupled through the radiation
field and, consequently, the hard X-ray spectral indices are  almost fixed 
at $\alpha\sim 1.1$.
In the soft-state solution,
most of the accretion energy is dissipated in
the disk and thus the disk is radiation pressure-dominated. The corona is
weak, characterized by lower temperature and density, and is no longer strongly
coupled with the disk through seed photons. The
emergent spectral energy comes mainly from the un-scattered intrinsic
disk radiation, peaking at UV and soft X-rays.
The soft states appear at only very high luminosity; 
while hard states appear at both high and low luminosities.
 A  state between soft and hard, composed of an inner soft-sate solution and 
an outer hard-state solution, is also possible for a moderately
 luminous system.
 Our model predicts that the spectra is
hard  from low-luminosity objects with $L\la 0.2L_{\rm Edd}$; 
While the spectra can be either
hard or soft  from high-luminosity objects with $L\ga 0.8L_{\rm Edd}$. 
 Spectra  from objects with luminosity 
 $0.2L_{\rm Edd}\la L\la 0.8L_{\rm Edd}$ 
can be hard, 
and can also be moderately soft, which is softer at higher luminosity.   

\acknowledgements{We would like to thank R. Matsumoto for providing part of
the simulation code. We are grateful to the anonymous referee for his/her 
detailed and helpful comments. 
BFL thanks for support by  Japan Society for the
Promotion of Science. This work is partially supported by
the Grants-in Aid of the Ministry of Education, Culture, Sports,
Science and Technology of Japan (P.01020 to B.F.L; 13640238 and 14079205 to  S.M.).}

\clearpage

\end{document}